\documentstyle[12pt] {article}
\begin {document}
\begin{center}
\vskip 1.5 truecm
\begin{Large}
{\bf Elastic $\rho '$ and $\phi$ Meson Photo- and Electroproduction with
Non-Resonant Background}\\
\end{Large}
\vspace{.5cm}
M.G.Ryskin and Yu.M.Shabelski \\
\vspace{.5cm} Petersburg Nuclear Physics Institute, \\
Gatchina, St.Petersburg 188350 Russia \\
\end{center}
\vspace{1cm}
\begin{abstract}
The backgrounds due to the direct diffractive dissociation of the photon
into the $\pi^+\pi^-$ as well as $K^+K^-$ pairs to the "elastic"
diffracti\-ve $\rho '$ and $\phi $ mesons production in electron-proton
collisions are calcula\-ted. It is shown that the role of background and
background-resonant interference contributions is very important in
experimental distribu\-ti\-on of $M_{\pi^+ \pi^-}$ in the $\rho '$ mass
region. The amplitude for the background process is proportional to the
$\pi$-meson - proton or $K$-meson - proton cross sections. Therefore,
describing the HERA data, one can estimate $\sigma (\pi p)$ and $\sigma
(Kp)$ at energies higher than the region of direct measure\-ments.
\end{abstract}
\vspace{3cm}

E-mail: $\;$ RYSKIN@thd.PNPI.SPB.RU  \\

E-mail: $\;$ SHABEL@vxdesy.desy.de \\

\newpage

\section{Introduction}
It was noted many years ago that the form of the $\rho$-meson peak is
distorted by the interference between resonant and non-resonant
$\pi^+\pi^-$ production. For the case of "elastic" $\rho^0$
photoproduction the effect was studied by P.S\"oding in \cite{so} and
S.Drell \cite{dr}. At high energies the main (and the only) source of
background is the Drell-Hiida-Deck process \cite{d} (see fig. 1). The
incoming photon fluctuates into the pion pair and then $\pi p$-elastic
scattering takes place. Thus the amplitude for the background may be
written in terms of the pion-proton cross section. It was demonstrated
\cite{zeus,H1} that the interference with some non-resonant background
is indeed needed to describe the distribution over the mass - $M$ of
$\pi^+\pi^-$ pair. M.Arneodo proposed \cite{ar} that this effect
can be used to estimate the value of $\sigma_{\pi p}$ from HERA data at
high energies, in the range which is not otherwise acceptable. Our
calculations \cite{RSh} show that really the data on $M$ distributions in
the case of $\pi^+\pi^-$ pair production can be described satisfactory
with the realistic value of $\sigma^{tot}_{\pi p} \sim$ 30 mb at energy
$s_{\pi p} \sim (2 \div 3) 10 ^3$ GeV$^2$.

To prove that this method of extraction of total $\pi p$ cross section
is really working, it should give the same values of
$\sigma^{tot}_{\pi p}$ for all resonances which decay into
$\pi^+\pi^-$ pair. So in the present paper we will consider the
mass distribution of produced $\pi^+\pi^-$ in the region of
$\rho '$ meson. The situation with $\pi^+\pi^-$ pair production in the
region of $\rho '$ mass is not simple because there are evidence
\cite{PDG} that here two different resonances exist, $\rho (1450)$
and $\rho (1700)$ and the "tail" from $\rho (770)$ production is not
negligibly small. However, if the amplitude of $\pi \pi$ scattering is
an analitycal function with good behaviour at infinity in $M_{\pi \pi}$
complex plane, the $M_{\pi \pi}$ distribution should be determined
totally by the singularities of elastic $\pi \pi$ amplitude, i.e. by
contributions of all resonances and the cut corresponding to the
$2\pi$ non-resonant production.

In Sect. 2 the formulae for the $2\pi$ background which are valid for
the DIS as well as for the photoproduction region are presented. The
expression differs slightly from the S\"oding's one as we take into
account the meson form factor and the fact that one meson propagator is
off-mass shell. We consider also the absorbtion correction comming from
the diagram where both mesons ( $\pi^+$ and $\pi^-$, or $K^+$ and $K^-$)
directly interact with the target proton. The role of the interference
in $\rho '$ photo- and electroproduction is discussed in Sect. 3.
We will also consider in Sect. 4 the case of $\phi$ photo- and
electroproduction which allows one, in principle, to extract the
value of $\sigma^{tot}_{Kp}$ at high energies.

\section{Production amplitudes}

The cross section of vector meson photo- and electroproduction may be
written as:
\begin{equation}
\frac{d\sigma^D}{dM^2dt}\; =\; \int d\Omega \;
|A_{r.}+A_{n.r.}|^2 \;,
\end{equation}
where $A_{r.}$ and $A_{n.r.}$ are the resonant and non-resonant parts of
the production amplitude, $D=L\, ,T$ for longitudinal and transverse photons,
$t = -q_t^2$ is the momentum transfered to the proton and
$d\Omega =d\phi dcos(\theta)$, where $\phi$ and $\theta$ are the azimutal
and polar angles between the $\pi^+$ and the proton direction in the
$2\pi$ rest frame.

\subsection{Amplitude for resonant production}

We will use the simple phenomenological parametrization of the
production amplitude because our main aim is the discussion of the
interference between resonant and non-resonant contributions. So the
amplitude for resonant process $\gamma p \to V p$ :
\begin{equation}
A_V\; =\; \sqrt{\sigma_V}e^{-b_V q^2_t/2}\frac{\sqrt{M_0\Gamma}}
{M^2-M^2_0+iM_0\Gamma}\frac{Y^D(\theta ,\phi)}{\sqrt{\pi}} \;,
\end{equation}
where $M_0$ and $\Gamma = \Gamma_0$ are the mass and the width of vector
meson; $b_V$ is the $t$-slope of the "elastic" $V$ production cross
section $\sigma_V \equiv d\sigma(\gamma p \to V p)/dt$ (at $t=0$) and
the functions $Y^D(\theta ,\phi),\; D = T, L$ describe the angular
distribution of the pions produced through the vector meson decay:
\begin{equation}
Y^0_1\; =\; \sqrt{\frac 3{4\pi}}cos\theta \;,
\end{equation}
\begin{equation}
Y^{\pm 1}_1\; =\; \sqrt{\frac 3{8\pi}}sin\theta\cdot e^{\pm i\phi} \;.
\end{equation}
Note that for transverse photons with polarization vector $\vec e$ one
has to replace the last factor $e^{\pm i\phi}$ in eq. (4) by the scalar
product $(\vec e\cdot\vec n)$, where $\vec n$ is the unit vector in the
pion transverse momentum direction.

\subsection{Amplitude for non-resonant production}

The amplitude for the non-resonant process $\gamma p \to \pi^+\pi^-p$ is:
\begin{equation}
A_{n.r.}\; =\;\sigma_{\pi p}F_\pi (Q^2)e^{bt/2}\frac{\sqrt{\alpha}}
{\sqrt{16\pi^3}} B^D\sqrt{z(1-z)\left|\frac{dz}{dM^2}\right|
\left(\frac{M^2}4-m^2_\pi\right)|cos\theta|} \;,
\end{equation}
where $b$ is the $t$-slope of the elastic $\pi p$ cross section,
$F_{\pi} (Q^2)$ is the pion electromagnetic form factor
($Q^2=|q^2_\gamma| > 0$ is the virtuality of the incoming photon),
$\alpha= 1/137$ is the electromagnetic coupling constant and $z$ -- the
photon momentum fraction carried by the $\pi^-$ -meson; $\sigma_{\pi p}$
is the total pion-proton cross section.

The factor $B^D$ is equal to
\begin{equation}
B^D\; =\; \frac{(e^D_\mu\cdot k_{\mu -})f(k'^2_-)}{z(1-z)Q^2+m^2_\pi+k^2_{t-}}
-\frac{(e^D_\mu\cdot k_{\mu +})f(k'^2_+)}{z(1-z)Q^2+m^2_\pi+k^2_{t+}}
\end{equation}
 For longitudinal photons the products $(e^L_\mu\cdot k_{\mu \pm})$ are:
 $(e^L_\mu\cdot k_{\mu -})=z\sqrt{Q^2}$ and
$(e^L_\mu\cdot k_{\mu +})=(1-z)\sqrt{Q^2}$, while for the transverse
photons we may put (after averaging) $e^T_\mu\cdot e^T_\nu =\frac
 12\delta^T_{\mu\nu}$.

Expressions (5) and (6) are the result of straitforward calculation of
the Feynman diagram fig. 1. The first term in (6) comes from the graph
fig. 1 (in which the Pomeron couples to the $\pi^+$) and the second one
reflects the contribution originated by the $\pi^- p$ interaction.  The
negative sign of $\pi^-$ electric charge leads to the minus sign of the
second term. We omit here the phases of the amplitudes. In fact, the
common phase is inessential for the cross section, and we assume that the
relative phase between $A_{r.}$ and $A_{n.r.}$ is small (equal to zero)
as in both cases the phase is generated by the same 'Pomeron'
\footnote{Better to say -- 'vacuum singularity'.} exchange.

The form factor $f(k'^2)$ is written to account for the virtuality
($k'^2\neq m^2_\pi$) of the t-channel (vertical solid line in fig. 1)
pion.  As in fig. 1 we do not deal with pure elastic pion-proton
scattering, the amplitude may be slightly suppressed by the fact that
the incoming pion is off-mass shell. To estimate this suppression we
include the form factor (chosen in the pole form)
\begin{equation}
f(k'^2)=1/(1 + k'^2/m'^2) \;.
\end{equation}
The same pole form was used for $F_{\pi} (Q^2)=1/(1 + Q^2/m^2_{\rho})$.
In the last case the parameter $m_{\rho}$ is the mass of the
$\rho$-meson, but the value of $m'$ (in $f(k'^2)$) is expected to be
larger. In the case of $\pi^+ \pi^-$ production it should be of the order
of mass of the next resonance from Regge $\pi$-meson trajectory; i.e.
it should be the mass of $\pi (1300)$ or $b_1 (1235)$. Thus we put
$m'^2=1.5$ GeV$^2$.

Finaly we have to define $k'^2_\pm$ and $k_{t\pm}$.
\begin{equation}
\vec k_{t-}=-\vec K_t+z\vec q_t\;\;\;\;\;\;\;
\vec k_{t+}=\vec K_t+(1-z)\vec q_t
\end{equation}
and
\begin{equation}
k'^2_-=\frac{z(1-z)Q^2+m^2_\pi+k^2_{t-}}{z},\;\;\;\;\;\;
k'^2_+=\frac{z(1-z)Q^2+m^2_\pi+k^2_{t+}}{1-z} \;.
\end{equation}
In these notations
$$M^2=\frac{K^2_t+m^2_\pi}{z(1-z)},\;\;\;\;
\;\;\;\; dM^2/dz=(2z-1)\frac{K^2_t+m^2_\pi}{z^2(1-z)^2}$$
and $ z=\frac 12\pm\sqrt{1/4-(K^2_t+m^2_\pi)/M^2}$ with the pion
transverse (with respect to the proton direction) momentum $\vec K_t$
(in the $2\pi$ rest frame) given by expression
$K^2_t=(M^2/4 - m^2_\pi)sin^2\theta$. Note that the positive values of
$cos\theta$ correspond to $z \geq 1/2$ while the negative ones
$cos\theta < 0$ correspond to $z \leq 1/2$.

\subsection{Absorptive correction}

To account for the screening correction we have to consider the diagram
fig. 2, where both pions interact directly with the target. Note that all
the rescatterings of one pion (say $\pi^+$ in fig. 1) are already included
into the $\pi p$ elastic amplitude. The result may be written in form of
eq. (5) with the new factor $\tilde{B}^D$ instead of the old one
$B^D=B^D(\vec K_t,\vec q)$:
\begin{equation}
\tilde{B}^D\; = \; B^D(\vec K_t,\vec q)-\int C\frac{\sigma_{\pi p}
e^{-bl^2_t}} {16\pi^2}B^D(\vec K_t-z\vec l_t,\vec q)d^2l_t
\end{equation}
where the second term is the absorptive correction (fig. 2) and $l_\mu$
is the momentum transfered along the 'Pomeron' loop. The factor $C>1$
reflects the contribution of the enhacement graphs with the diffractive
exitation of the target proton in intermediate state. In accordance with
the HERA data \cite{H2}, where the cross section of "inelastic" (i.e.
with the proton diffracted) $\rho$ photoproduction was estimated as
$\sigma^{inel}\simeq 0.5 \sigma^{el}$ we choose $C=1.5 \pm 0.2$.

\section{Elastic $\rho '$ photo- and electroproduction cross section}

The situation with resonance $\rho '$ is not clear. The enhancement
in the four pion invariant mass distribution at $\sim 1.6$ GeV which
was considered as $\rho '(1600)$ particle now can be considered
\cite{PDG} as two interfering resonances, $\rho (1450)$ and $\rho (1700)$
($r1$ and $r2$ in the formulae below) with masses 1.465 GeV and 1.70 GeV
and widths 0.31 GeV and 0.235 GeV, respectively. However the presented
formulae are valid for one-resonant production. The relative phase of the
amplitudes of these resonances production is practically unknown. For
these reason we will present the calculated results for two cases, one
resonance with $M = 1.57$ GeV and $\Gamma = 180$ MeV \cite{H1a}, using
eq. (1), and two-resonances with the masses and widths presented above
and with zero relative phase. In the last case we will use expression
\begin{equation}
\frac{d\sigma^D}{dM^2dt}\; =\; \int d\Omega \;
|- A_{r1} - A_{r2} + A_{n.r.}|^2 \;
\end{equation}
instead of eq. (1) and in agreement with paper \cite{Abe} we use the
negative sign of the resonant contribution that can be connected with
the negative signs of $r1, r2 \to \pi \pi$ decay amplitude. The main aim
of our calculations is to demonstrate the role of background
contribution to the resulting $M_{2\pi}$ distribution.

The results at $Q^2=0$ (photoproduction) for the cases of one-resonance
and two-resonances production are shown in figs. 3a and 3b. Both slope
parameters, $b_{\rho}$ in eq.(2) and $b$ in eq.(5) were assumed to be
equal to 10 GeV$^{-2}$. The dashed curves correspond to the resonant
production contributions. They are normalized to the experimental
$\rho '$ to $\rho$ ratio in the two-pion final state (i.e.  the product
of cross section ratio and $\rho ' \to \pi^+\pi^-$ branching) that is
equal to $0.0134 \pm 0.0023$ at $Q^2=0$ \cite{Abe} and we assume that in
the case of two-resonances production their production cross sections are
equal. The values of two-pion branching ratios for these states were
assumed to be 0.07 and 0.25, respectively \cite{PDG}.

The contributions of non-resonant production (dotted curves) and
its interference with resonant one (dash-dotted curves) depend on the
form factor $f(k'^2)$, the screening corrections and the value of
$\sigma_{\pi p}$. Here the form factor was used in the form eq.(7) with
$m^{'2}$ = 1.5 GeV$^2$, screening correction in the form eq.(10) and the
value of $\sigma_{\pi p} = 31$ mb were used \footnote{ZEUS analysis 
\cite{sig} of
$\pi^+ \pi^-$ production in the $\rho$ mass region of elastic
photoproduction.}. The sum of all three contributions are presented by
solid curves. One can see that the contributions of non-resonant and
interference production dominate for the mass region of $\rho(1450)$ and
the experimental enhancement in the two-pion mass distribution here is
connected mainly with background and interference contributions as well
as with possible "tails" from $\rho(770)$ and $\rho(1700)$ production.
This enhancement should be sensitive to the value of $\sigma_{\pi p}$.

To demonstrate the role of the relative sign of resonant and
non-resonant (background) production we present in figs. 3c and 3d the
calculations with the positive sign of resonant contribution in eq.
(11). Now we present in fig. 3d the sum of two resonances and their
interference by dashed curve, the sum of all resonant-background
contributions by dash-dotted curve and background contribution by
dotted curve. One can see that the shape of the summary curves (solid
ones) are changed drastically in comparison with the cases of figs. 3a
and 3b, especially in the case of two-resonant production.

We have considered also the process of $\pi^+ \pi^-$ production in
the $\rho '$ mass region with accounting the "tail" of $\rho (770)$
contribution, i.e. using the expression
\begin{equation}
\frac{d\sigma^D}{dM^2dt}\; =\; \int d\Omega \;
|A_{\rho} - A_{r1} - A_{r2} + A_{n.r.}|^2 \;.
\end{equation}
It is interesting that the contributions of $\rho (770)$
Breit-Wigner peak and its interference with background practically
cancel at the used parameters (on the level of $\sim$ 90\%) the
background contribution shown in figs. 3. It means that
$\pi^+ \pi^-$ production in the mass region, say 1.2-1.9 GeV is
determined practically only by one or two $\rho '$ resonances
\footnote{As the accuracy of this cancellation depends crucially
on the values of used parameters, one has to check the "background"
contribution experimentally, say in the region of $M \sim 1$ GeV.}.

At very large $Q^2$ the background amplitude eq.(5) becomes negligible
as, even without the additional form factor (i.e. at $f(k'^2)\equiv 1$),
the non-resonance cross section falls down as $1/Q^8$ \footnote{In the
amplitude $A_{n.r.}$ one factor $1/Q^2$ comes from the electromagnetic
form factor $F_\pi(Q^2)$ and another one -- from the pion propagator
(term - $z(1-z)Q^2$ in the denominator of $B^D$ (see eq.(6)).}.

The results of our calculations of $d\sigma /dM$ for $Q^2$ = 10 GeV$^2$
using eq. (1) for one $\rho '$ resonance and eq. (11) for two
resonances with form factor, eq.(7), $\sigma_{\pi p} = 31$ mb and with
screening correction are presented in figs. 4a and 4b, respectively.
They are normalized to the experimental \cite{H1a} $\rho '$ to $\rho$
ratios ($0.36 \pm 0.07 \pm 0.11$ for $\pi^+ \pi^- \pi^+ \pi^-$ channel)
which were assumed to be equal to 0.8 for one $\rho(1570)$ resonance and
to 0.4 at for both states, $\rho(1450)$ and $\rho(1700)$. The pure
resonant contributions (the sum of two resonances and their interference
in the case of fig. 4b) are shown by dashed curves (we assume that the
longitudinal contribution to the resonant production is equal to the
transverse one) and one can see that in both cases resonant production
dominate.

The contribution of $\rho (770)$ here is small. For
$M_{\pi \pi} \sim 1.4 $ GeV the $\rho(770)$ "tail" is 5-6 times
smaller than the non-resonant $\pi^+\pi^-$ background.

\section{Elastic $\phi$ photo- and electroproduction $\;\;\;$ cross
section}

In the case of $\phi$-meson photo- and electroproduction the corrections
to the pure resonant production should be small because the resonant
peak is very narrow. However their contribution should be investigated.

We will consider the channel $\phi \to K^+K^-$ with branching ratio
$49.1 \pm 0.6$\%. The absolute normalization of the $\phi$
photoproduction cross section can be obtained by integration of $K^+K^-$
mass distribution near the $\phi$ peak that should corresponds to the
value $\sigma_{\gamma p \to \phi p} = 0.96 \pm 0.19^{+21}_{-18} \mu$b
\cite{zeus2}. The value of $b_{\phi}$ parameter equal to 7 GeV$^{-2}$ in
eq. (2) for $t$-slope of the elastic $\phi$ photoproduction was used
in agreement with data of \cite{zeus2}. Here the form factor $f(k'^2)$,
was used in the form eq.(7) with $m^{'2}$ = 2 GeV$^2$, that corresponds
to the mass of the next resonance from Regge $K$-meson trajectory.
We take the value $b$ = 9 GeV$^{-2}$ for $t$-slope in elastic $Kp$ cross
section and $\sigma^{tot}_{Kp}$ = 24 mb for the total cross section of
$Kp$ interactions.

The results of calculations for the case of $K^+K^-$ photoproduction are
shown in fig. 5a. Of course, resonant contribution dominates in the
region of $\phi$ peak, however some interference contribution can be seen
at the right-hand side of the "tail". So in principle it is possible to
estimate total $Kp$ cross section, however for this problem it is
necessary to have very good identification of secondary kaons
\footnote{Now the main source of experimental background for $\gamma p
\to \phi p$ reaction is $\gamma p \to \rho p$ process \cite{zeus2}.}.

The results for $K^+K^-$ electroproduction cross section at $Q^2$ = 10
GeV$^2$ in the mass region of $\phi$ peak is shown in fig. 5b. Here we
assumed that the longitudinal contribution to the resonant production is
1.5 times larger than the transverse one and one can see that the
resonance production dominate.

\section{Conclusion}
We presented simple formulae for the background to elastic
$\rho '$-meson and $\phi$ photo- and electroproduction which account
for the absorptive correction and virtuality of the pion and kaon. The
role of resonance -- background interference is very significant for the
case of $\pi^+ \pi^-$ pair photoproduction in the mass region of
$\rho '$-meson. Due to some occacional reasons the "tail" of $\rho(770)$
is practically cancelled by background contribution.
At large $Q^2$, say at $Q^2\sim 10$ GeV$^2$, for the case of $\rho '$
production the $\rho(770)$ "tail" is rather small and the main
background comes from non-resonant $\pi^+\pi^-$ production. For the
case of $\phi$ production background is small enough.

We are grateful to M.Arneodo for stimulating discussions. The paper is
supported by INTAS grant 93-0079 and by Volkswagen Stiftung.

\newpage

\begin{center}
{\bf Figure captions}\\
\end{center}

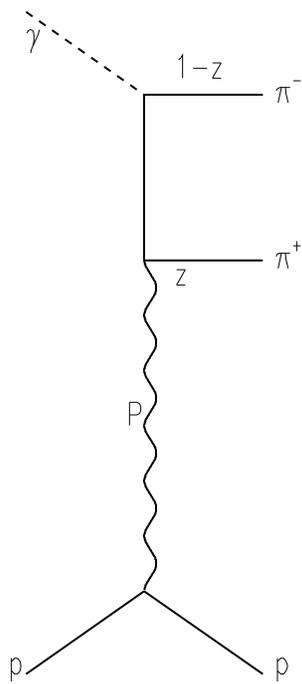
Fig. 1. Feynman diagram for the two pion photo-(electro)production.

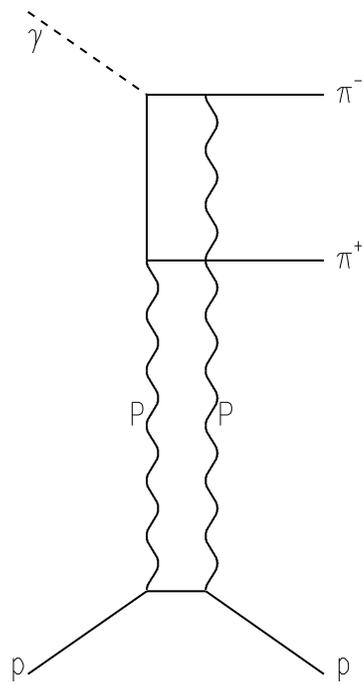
Fig. 2. Diagram for the absorbtive correction due to both pions (kaons)
rescattering.

Fig. 3. Resonant (Breit-Wigner, dashed curves), non-resonant
backgro\-und (dotted curves) and their interference (dash-dotted curves)
contributions to $\gamma p \to \pi^+ \pi^- p$ photoproduction process
for the cases of one (a,c) and two (b,d) resonances in the $\rho '$ mass
region. The cases of figs. a, b and c,d have different signs of the
interference contributions, see text. The sums of all contributions are
shown by solid curves.

Fig. 4. Resonant (Breit-Wigner, dashed curves), non-resonant backgro\-und
(dotted curves) and their interference (dash-dotted curves) contributions
to $\gamma p \to \pi^+ \pi^- p$ electroproduction process at
$Q^2$ = 10 GeV$^2$ for the cases of one (a) and two (b) resonances in the
$\rho '$ mass region. The sums of all contributions are shown by solid
curves.

Fig. 5. Resonant (Breit-Wigner, dashed curve), non-resonant background
(dotted curve) and their interference (dash-dotted curve) contributions
to $\gamma p \to K^+ K^- p$ photoproduction process. The sums of all
contributions are shown by solid curve (a). The case of elastic $K^+K^-$
electroproduction at $Q^2$ = 10 GeV$^2$ (solid curve); the dashed curve
presents the resonant contribution (b).

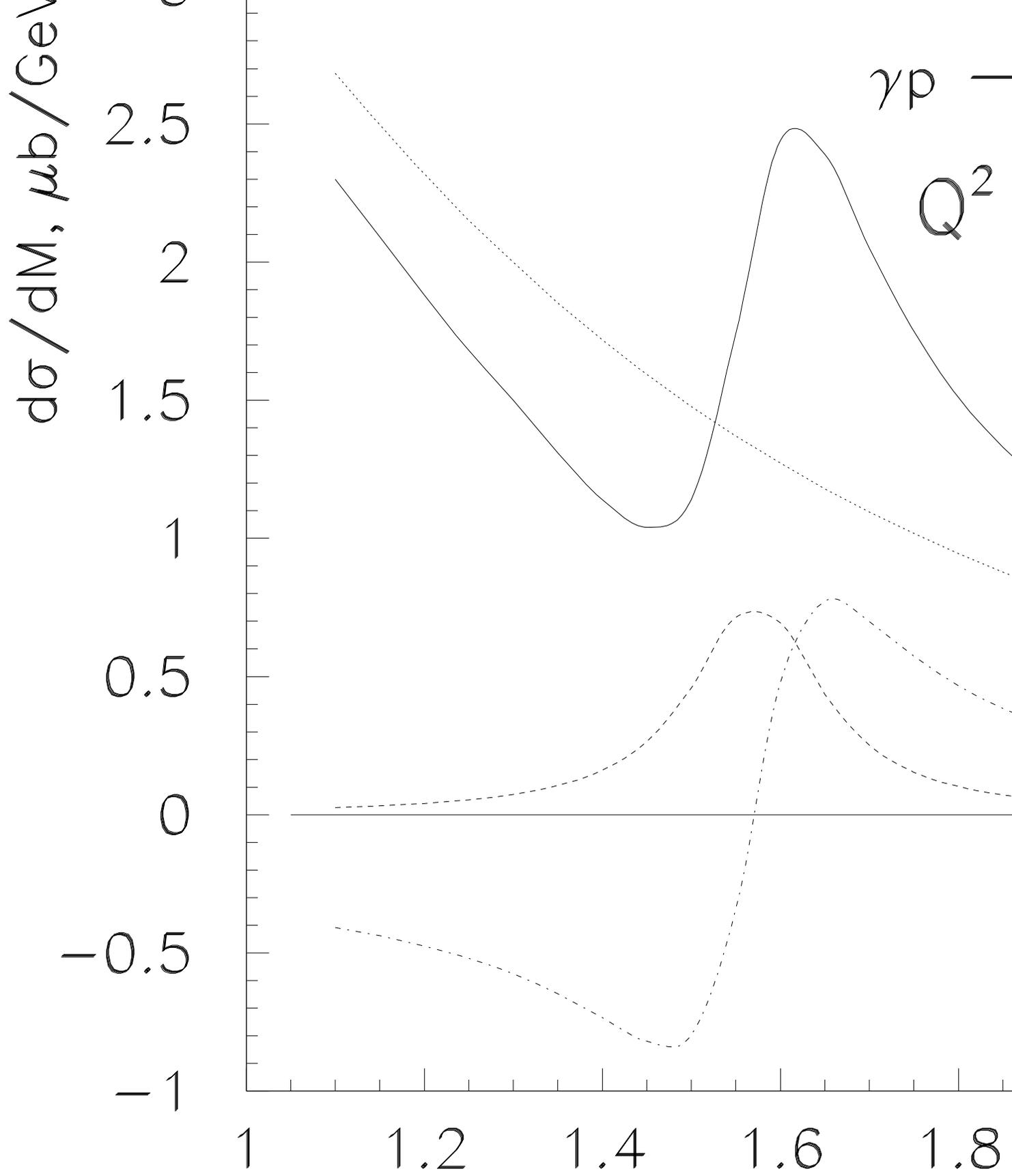
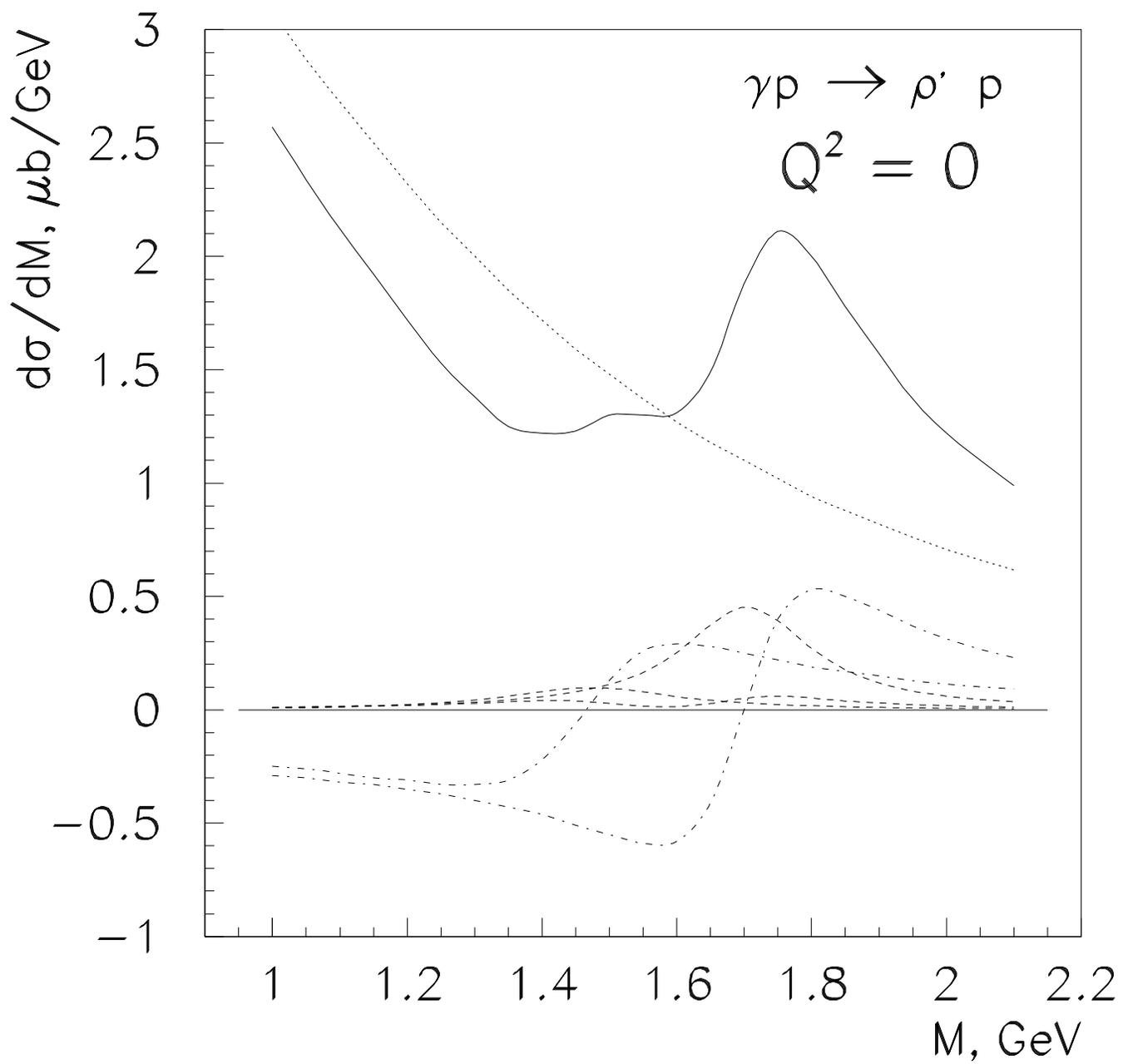
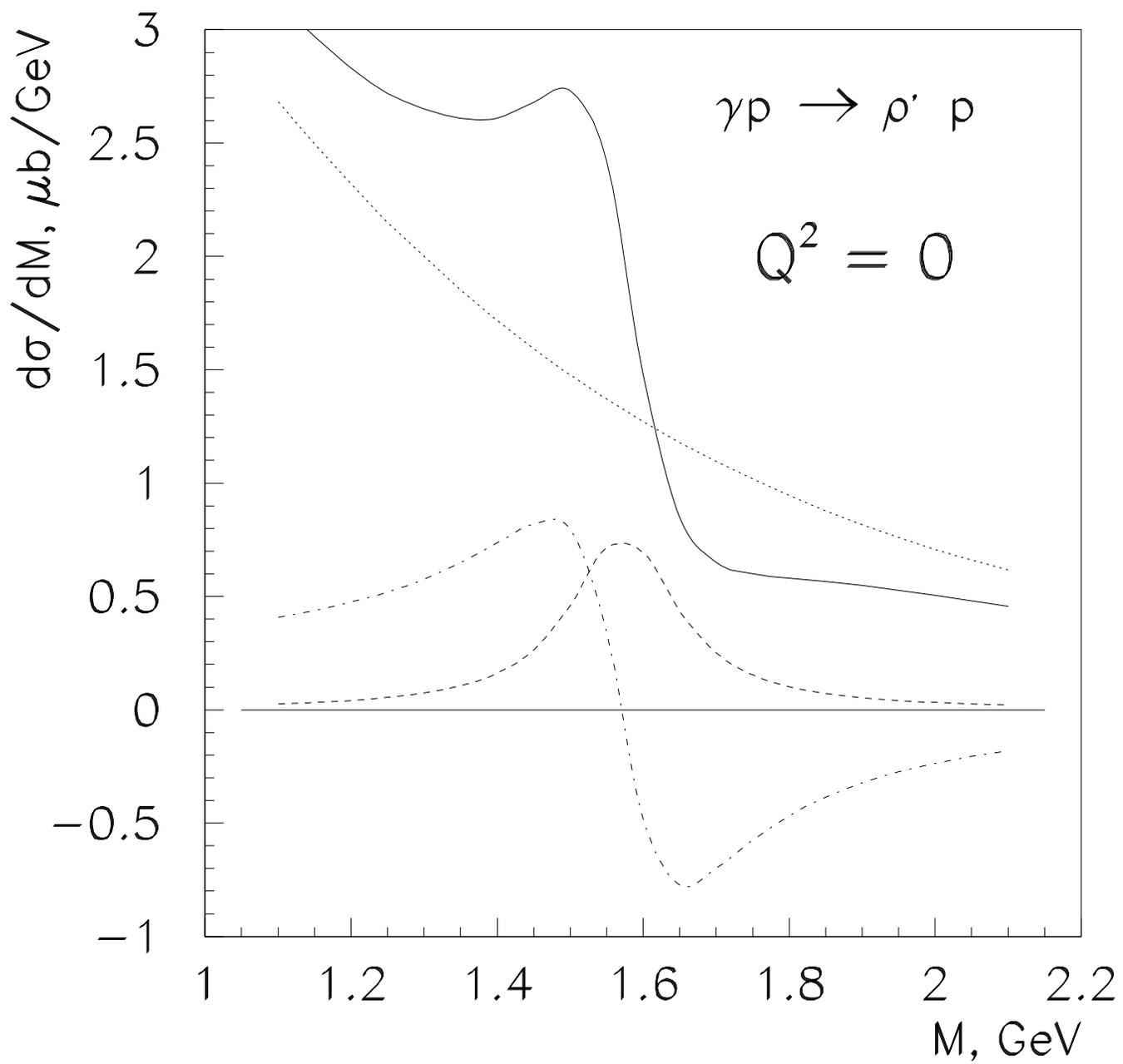
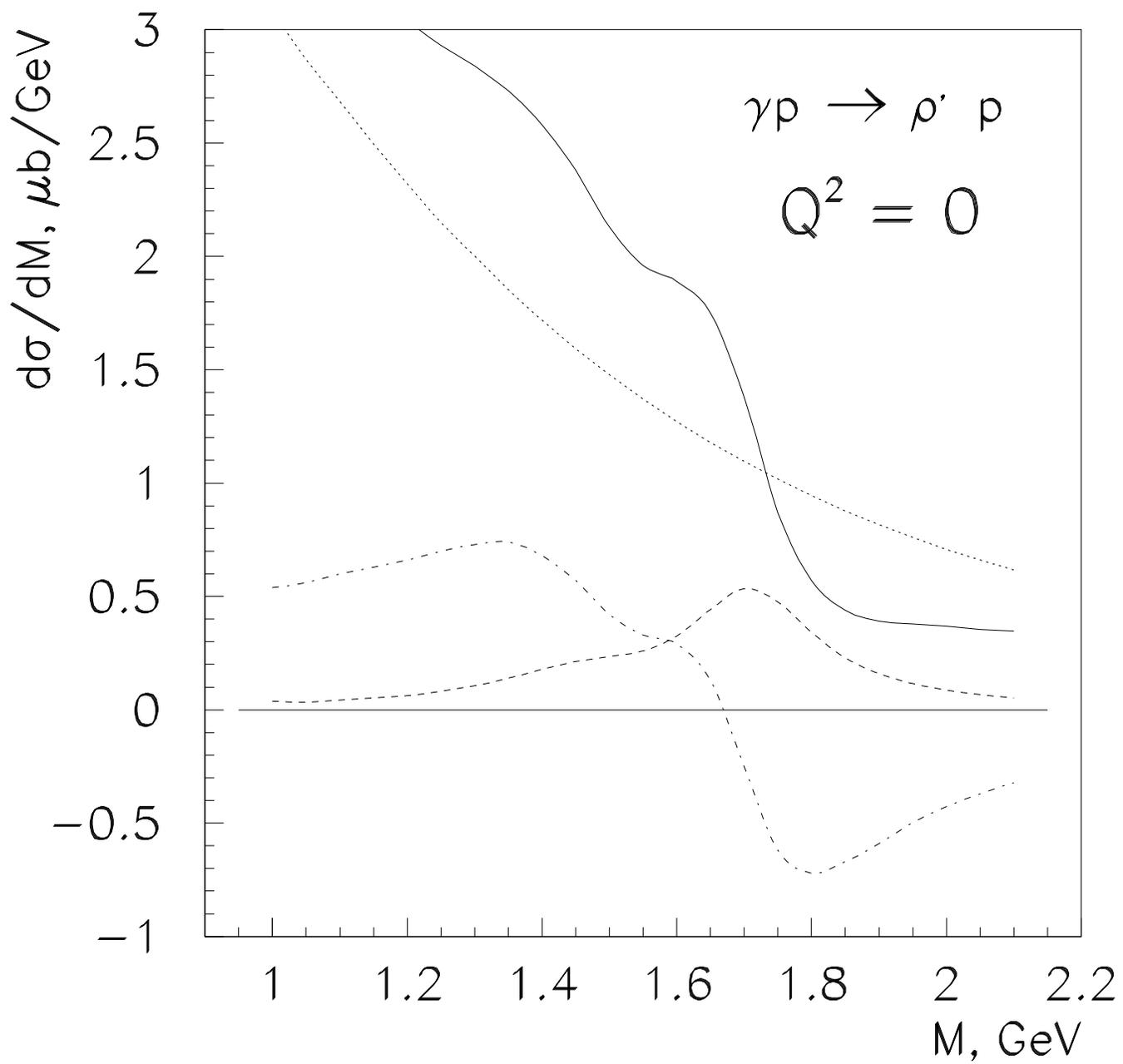
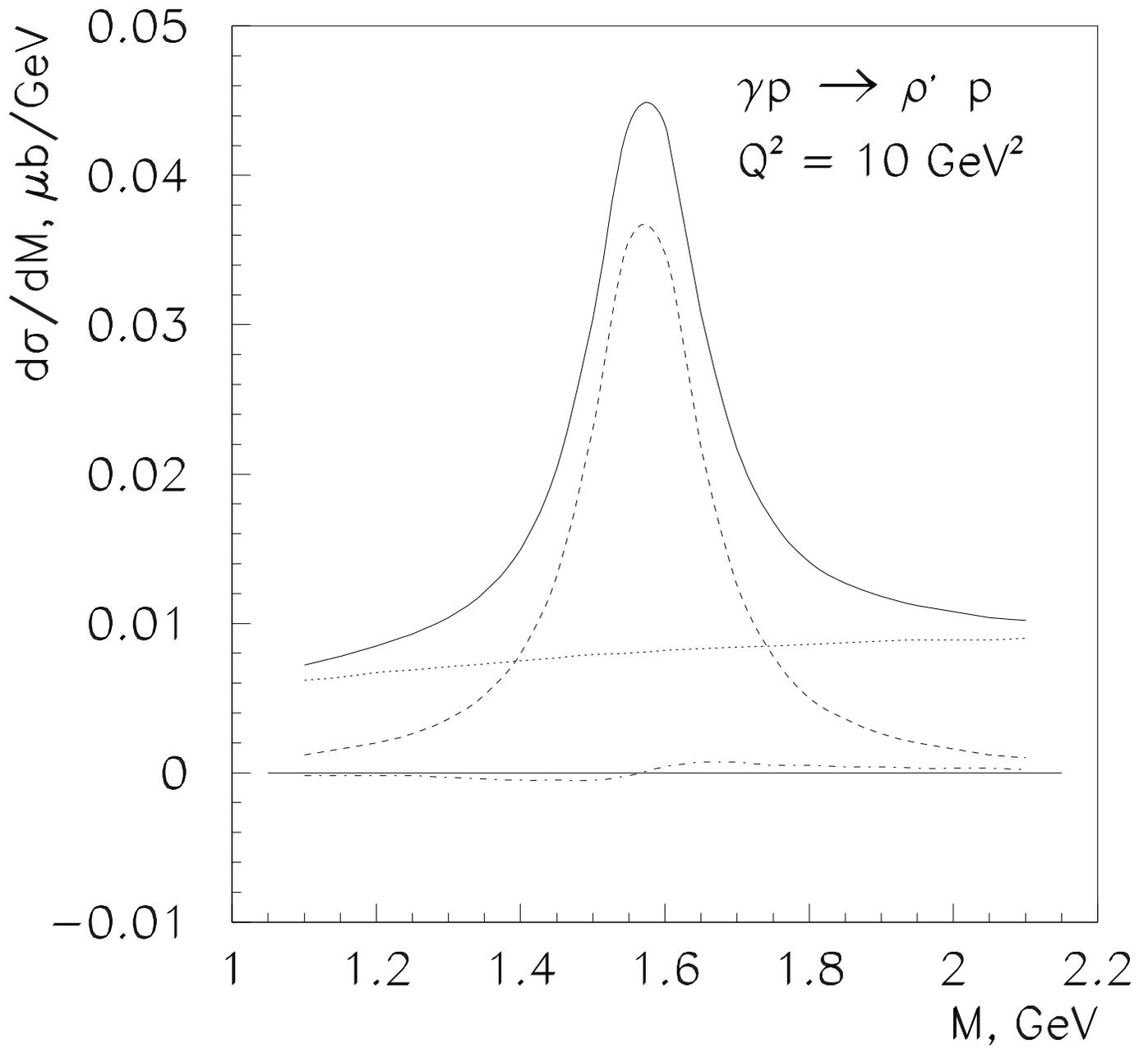
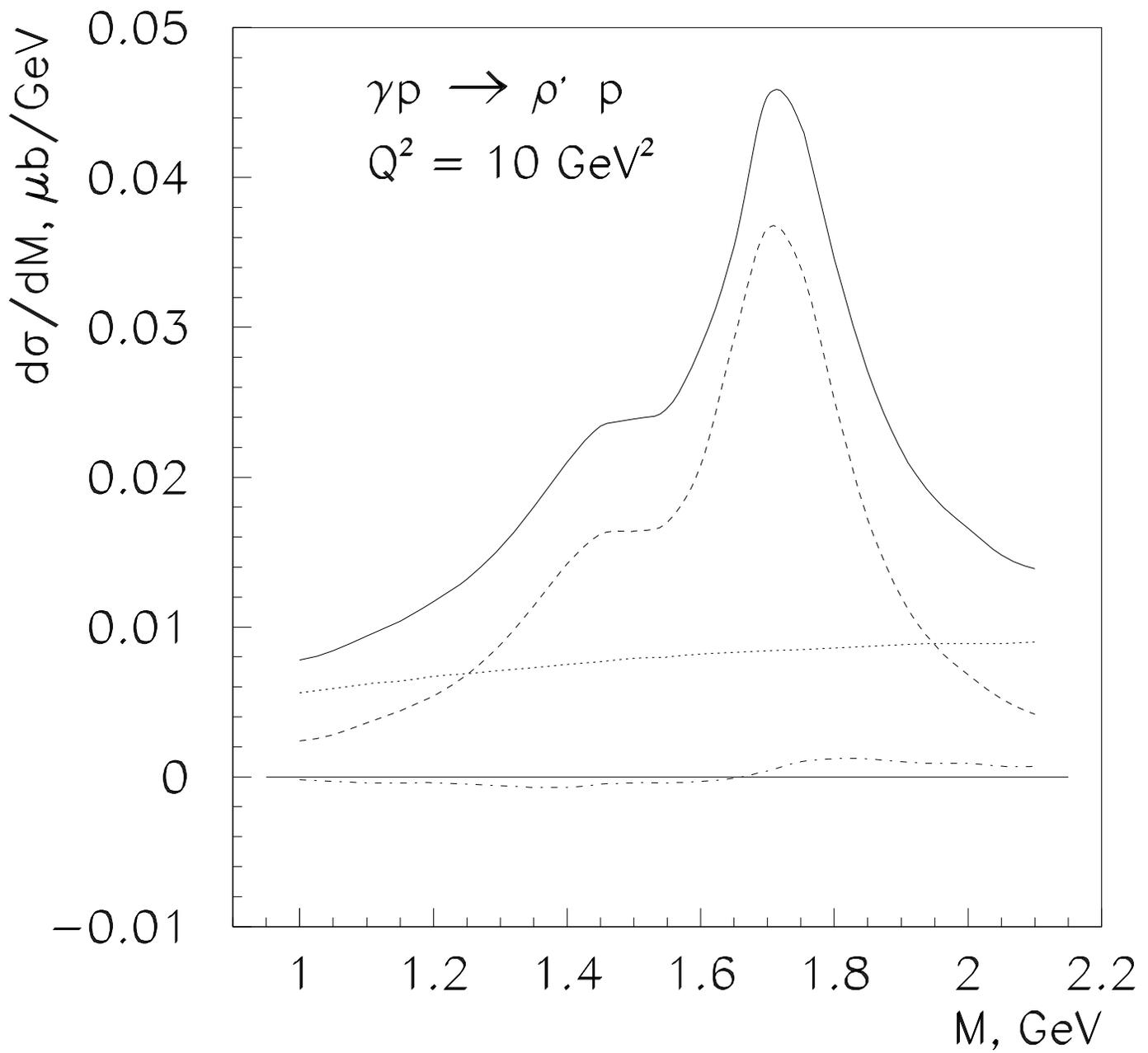
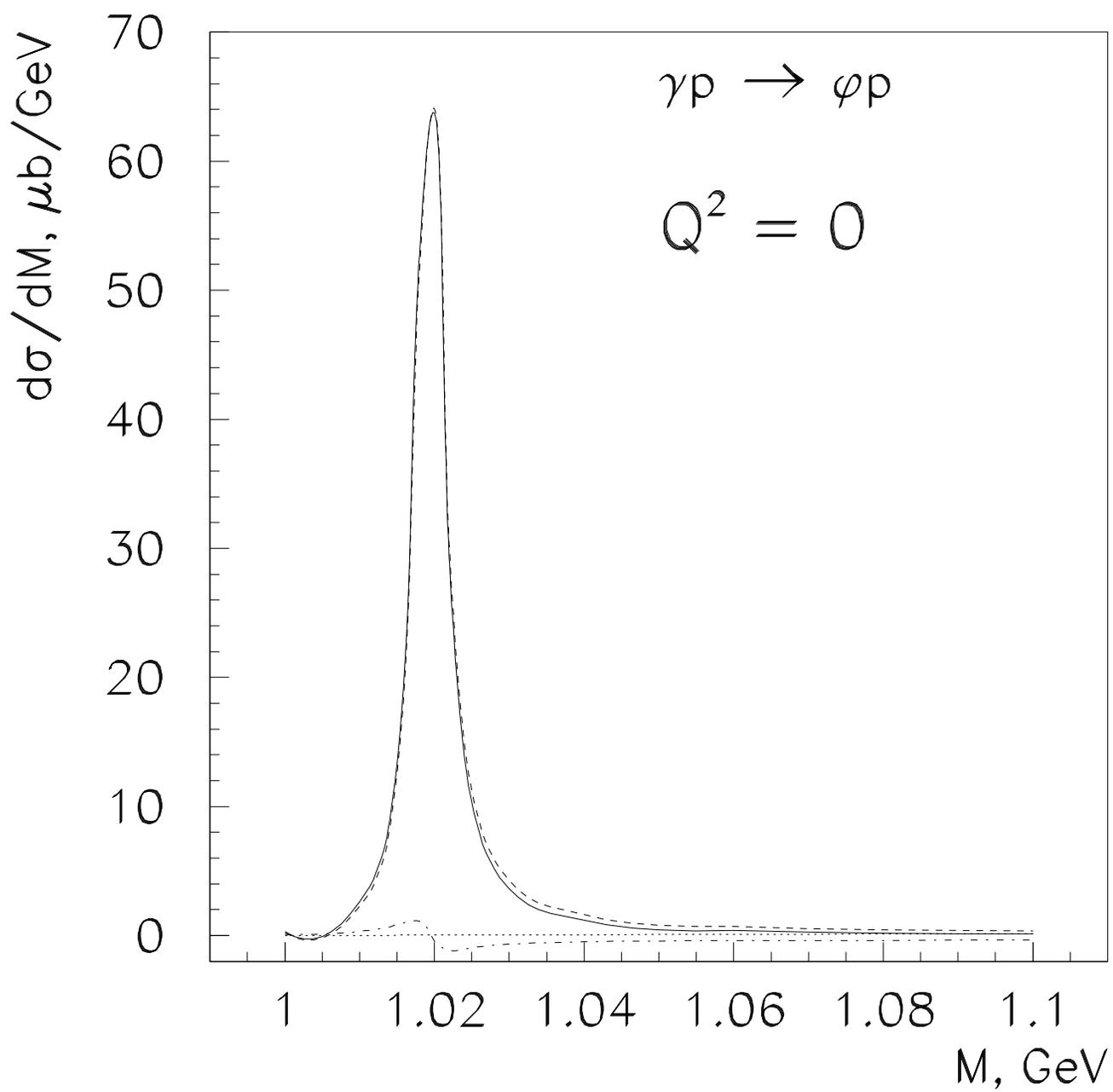
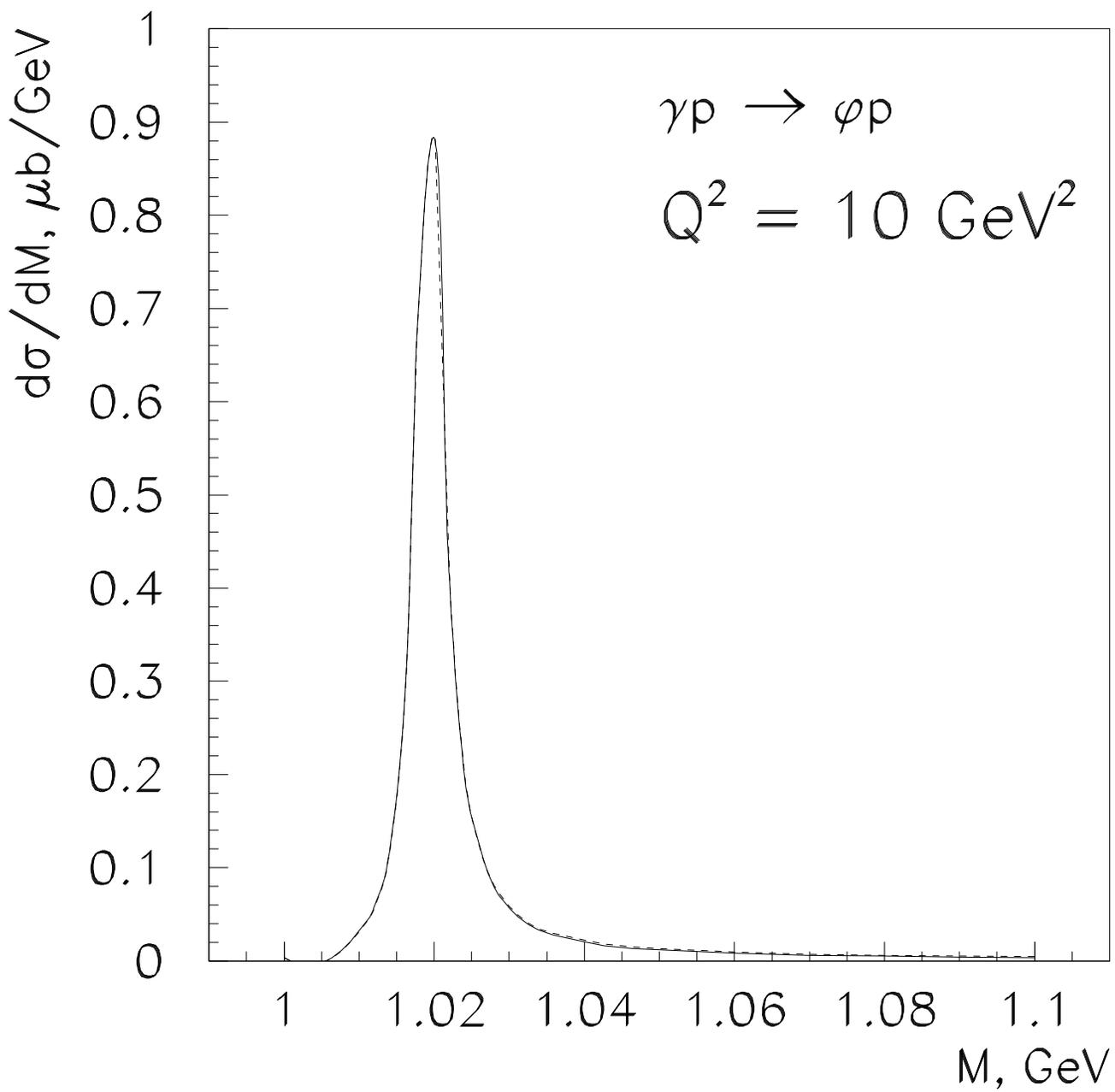
\end{document}